\documentclass[aps,prb,superscriptaddress]{revtex4-2}
\usepackage{amsmath}
\DeclareMathOperator{\arccosh}{arccosh}
\usepackage{amssymb}
\usepackage{amsthm}
\usepackage{graphicx}
\usepackage{bbold}

\begin{document}


\title{Hidden Bethe states in a partially integrable  model}

\author{Zhao Zhang}
\email{zz8py@virginia.edu}
\affiliation{SISSA and INFN, Sezione di Trieste, via Bonomea 265, I-34136, Trieste, Italy}
\affiliation{Tsung-Dao Lee Institute, Shanghai Jiao Tong University, Shanghai 200240, China}

\author{Giuseppe Mussardo}
\email{mussardo@sissa.it}
\affiliation{SISSA and INFN, Sezione di Trieste, via Bonomea 265, I-34136, Trieste, Italy}

\date{\today}

\begin{abstract}
We present a one-dimensional multi-component model, known to be partially integrable when restricted to the subspaces made of only two components. By constructing fully anti-symmetrized bases, we find  integrable excited eigenstates corresponding to the totally anti-symmetric irreducible representation of the permutation operator in the otherwise non-integrable subspaces. We establish rigorously the breakdown of integrability in those subspaces by showing explicitly the violation of the Yang-Baxter's equation. We further solve the constraints from Yang-Baxter's equation to find exceptional momenta that allows Bethe Ansatz solutions of solitonic bound states. These integrable eigenstates have distinct dynamical consequence from the embedded integrable subspaces previously known, as they do not span their separate Krylov subspaces, and a generic initial state can partly overlap with them and therefore have slow thermalization. However, this novel form of weak ergodicity breaking contrasts that of quantum many-body scars in that the integrable eigenstates involved do not have necessarily low entanglement. Our approach provides a complementary route to arrive at quantum many-body scars since, instead of solving towers of single mode excited states based on a solvable ground state in a non-integrable model, we identify  the integrable eigenstates that survive in a deformation of the Hamiltonian away from its integrable point.
\end{abstract}


\maketitle

\section{\label{sec:intro}Introduction}

Recent progress in cold atom experiments have made possible the simulation of a larger variety of strongly correlated systems~\cite{RevModPhys.80.885}. In particular, alkine-earth atoms such as $^{137}$Yb and $^{87}$Sr in optical atoms can be used to realize the Mott insulating phase of Fermi-Hubbard model, which can be described by the SU(N) Heisenberg Hamiltonian~\cite{Gorshkov:2010aa,Taie:2012aa,Pagano:2014aa,Scazza:2014aa,Zhang_2014}. This multi-component generalization of the spin-$\frac{1}{2}$ Heisenberg model was solved based on Yang's generalization ~\cite{yangprl,Sutherlandprl} of the Bethe Ansatz method developed by Sutherland~\cite{Sutherlandprb}. Using the same nested Bethe Ansatz method, Babelon, de Vega, and Viallet constructed an integrable $Z_{n+1} \times Z_{n+1}$ symmetric generalization of the XXZ model, from the solution of Yang-Baxter's equations~\cite{Babelon}. However, the anisotropy in this model was artificially designed so that the model is exactly solvable but not realistic to be realized in experiment. 

In this paper, we propose a minimal generalization of the spin-$\frac{1}{2}$ XXZ Hamiltonian, written in terms of permutation operators, whose SU(N) symmetry is explicitly broken to the symmetric group $\frak{S}_N$ by a diagonal term composed of the Cartan operators. As shown in Sutherland's original formulation of the isotropic model~\cite{Sutherlandprb}, such a multi-component spin model can be equivalently formulated in the language of one-dimensional hard-core bosonic particles with density-density interactions. The model proposed in this paper is in part motivated by a recent quantum system bearing a number theory  analogy~\cite{mussardo2020prime}, whose low energy effective theory can be seen as the current model in certain limits of its parameters. Some of the questions raised in the paper \cite{mussardo2020prime}, such as the the scaling of spectral gap from ground state, are partially answered in this paper.

It is worth to underline that, 
while the model proposed here is in general not integrable, the Krylov subspaces spanned by configurations involving only two components are nonetheless integrable, as was originally noted in a series of papers in the past ~\cite{Kiwata:1994vs,Sato:1995we,Sato:1996vm}. Hence, such a model falls  in the category of quasi-exactly solvable models, which has been studied since long in quantum mechanics~\cite{quasiesm}. In many-body systems, there is a class of models called frustration-free, whose zero energy ground state can be explicitly written as the common lowest energy eigenstate of each local operator in the Hamiltonian. Examples include the Majumdar-Ghosh model~\cite{Majumdar:1969aa}, AKLT model~\cite{PhysRevLett.59.799}, Motzkin and Fredkin spin chains~\cite{PhysRevLett.109.207202,Movassagh13278,Zhang5142,Salberger:2017aa,Salberger_2017,Zhang_2017} and,  in two dimensions, Kitaev's toric code model~\cite{Kitaev:2003aa}, to name a few. Many of these models boast solvable eigenstates beyond ground state as well. For instance, stabilizer code Hamiltonians consist of mutually commuting operators, so the entire spectrum is solvable. Despite the fact that neither Majumdar-Ghosh~\cite{Caspers:1982aa}, nor spin-1 AKLT model~\cite{Arovas:1989aa} satisfy this condition, they are known to have exact excited states. Recently these models have attracted a lot of attention for the experimental observation of anomalous dynamics in close related Rydberg atom experiments \cite{Bernien:2017aa} that interpolates between ergodicity breaking and thermalization behaviours. Such a dynamics has been attributed to a large overlap of the initial state with a tower of numerically solved eigenstates with equally spaced energies that violate the eigenstate thermalization hypothesis (ETH) in the so-called PXP Hamiltonian~\cite{PhysRevB.69.075106,PhysRevA.86.041601}, used to model the Rydberg blockade experiment~\cite{Turner:2018aa}. The discovery of this new form of weak ergodicity breaking has drawn more attention to analytical efforts for  solving exact excited states in non-integrable models~\cite{PhysRevLett.123.147201,PhysRevB.98.235155,PhysRevB.101.024306,PhysRevB.101.174308,PhysRevLett.124.180604,moudgalya2020large,PhysRevLett.119.030601,PhysRevResearch.1.033144,PhysRevB.101.241111,surace2020weakergodicitybreaking,SciPostPhys.3.6.043,PhysRevLett.122.220603,PhysRevB.105.035146}. The tremendous success achieved in searching these quantum many-body scar (QMBS) states is largely due to the approach which examines the sparsity of entanglement spectrum for numerically exactly diagonalized states~\cite{PhysRevB.98.235155}. In this way, these states can be analytically represented in terms of matrix product states~\cite{moudgalya2020large} making use of a reverse engineering. Furthermore, a large portion of these exact eigenstates share the common features of having energy expressed in terms of either integer or rational numbers, and they can be identified using single mode approximation (SMA) with $\pi$-momenta~\cite{moudgalya2020large}.
Some of these features observed in individual examples have found a more general explanation in the unified framework of spectrum generating algebra~\cite{PhysRevB.101.195131}. Concerning the partially integrability of the model proposed in this paper, it is useful to say that it  does not lead to the slow dynamics of weak ergodicity breaking, as the integrable subspaces studied before~\cite{Kiwata:1994vs,Sato:1995we,Sato:1996vm}. 
Indeed, depending on the subspace to which  an initial state belongs to, the system either thermalizes according to ETH, or exhibits strong ETH-violation due to partial integrability and behaves as a generalized Gibbs ensemble.

The plethora of frustration-free models (and their success in providing insights in understanding various aspects of many-body systems and quantum computing through rigorous mathematical theorems~\cite{Bravyi:2011wg,PhysRevLett.116.097202,Gosset:2016wx,Lemm:2019uz,10.1145/3357713.3384292,https://doi.org/10.48550/arxiv.2102.07209}), gives naturally rises to the question: to what extent can a frustration-free Hamiltonian interpolate between the situation in which only the ground state is known, and a stabilizer code that provides the entire solvable spectrum? The examples of exact QMBS in AKLT models partially answered this question, in that the additional eigenstates are of momenta $\pi$, allowing the diffractive scattering to be canceled out in a momentum eigenstate~\cite{PhysRevB.98.235155}. In this work, we will extend this idea to non-diffractive scattering with other momenta using Bethe Ansatz, in a non-integrable multi-component model with a ``weaker'' frustration. Such a strategy of easing frustration by enlarging local degrees of freedom has been employed by models with SU(N)-singlet simplex solid ground states \cite{PhysRevB.77.104404}, in the form of geometric frustration, which has also one-dimensional cousin models with longer range interaction, such as SU(3) spin chains with trimer and valance bond solid (VBS) ground states \cite{PhysRevB.75.060401}, and SU(N) VBS \cite{PhysRevB.75.184441}. Here, instead of designing a Hamiltonian as sum of projectors pinning down a desired ground state, we work with a minimal multi-component generalization to the spin-$\frac{1}{2}$ XXZ spin chain with 2-local interaction which violates the Yang-Baxter's equation (YBE) and therefore is not integrable. We show that the reduced frustration with larger local Hilbert space allows us to have additional Bethe Ansatz integrable eigenstates in any generic non-integrable Krylov subspace. Unlike the partial integrability studied before, these integrable eigenstates now bear a  new form of weak ergodicity breaking, as a generic initial state in these subspaces will have overlap with both integrable and non-integrable eigenstates.

Entanglement entropy and its scaling has been a major tool in probing many-body systems with strong correlations, see \cite{Laflorencie:2016aa} for a review. For one-dimensional systems, Hastings have given a rigorous proof that the ground state entanglement entropy scaling with system size is bounded by a constant for gapped systems \cite{Hastings:2007aa}. Meanwhile, people have been searching for area-law violating grounds states in gapless systems with logarithmic \cite{PhysRevLett.109.207202,Salberger:2017aa}, power law \cite{Movassagh13278}, and even linear scaling \cite{Zhang5142,Salberger_2017,Zhang_2017,Vitagliano:2010aa,Ramirez:2014aa}. In the study of quantum many-body scars, the sparsity of entanglement spectrum and the sub-volume law scaling of entanglement entropy has been used to show that the exact excited states violate strong eigenstate thermalization hypothesis \cite{PhysRevB.98.235155}. With the enlarged local Hilbert space, the entanglement entropy of the ground state of our model scales not only with the system size, but with the dimensionality of local Hilbert space as well. By rigorously establishing a volume law scaling of the ground state in the extreme frustration-free case of number of components equal to the system size (which also accounts for the additional contribution to the entanglement of integrable excited states from the antisymmetrized basis in addition to that from the corresponding Bethe Ansatz states in the spin-$\frac{1}{2}$ XXZ model\cite{Alba_2009,Chen_2013}), we show rigorously that the entanglement entropy scaling of our integrable eigenstates interpolates between area-law (which violates  entanglement entropy of corresponding excited states in XXZ chain) and volume law scaling of the totally antisymmetric bases.

The paper is organized as follows. In Sec.~\ref{sec:hamsym}, we introduce the model and discuss its symmetries, which we will use later for achieving the diagonalization of its Hamiltonian. Sec.~\ref{sec:diag} is rather articulated: in \ref{sec:sing} we first construct a proper basis to diagonalize the integrable subspaces of the Hilbert space; in \ref{sec:2bdy} we  illustrate the previous procedure in terms of an elementary approach for  diagonalizing the two-body case; in \ref{sec:mb}, using the general coordinate Bethe Ansatz, we  establish the embedding of the spectrum of spin-$\frac{1}{2}$ Heisenberg in our model. In Sec.~  \ref{sec:multi}, we rigorously prove the that a generic sector with multiple identical particles is non-integrable unless the anisotropy is turned off, and we use this result to shed more light on the reason of the partial integrability of the sectors discussed in the previous section. We also provide an explanation to the $\pi$-momenta QMBSs emerging from the condition to satisfy YBE with a generic anisotropy. In Sect.~\ref{sec:entropy}, we address 
the case of a large number of components (this makes the ground state frustration free) 
and we show analytically that the entanglement entropy of the ground state scales linearly in the thermodynamic limit. We also derive a rigorous decomposition of entanglement entropy contribution from the Bethe Ansatz wave-function and the antisymmetrized bases. In Sec.~\ref{sec:summ}, we draw our conclusions, also pointing out few interesting directions worth exploring in the future. The paper has also three appendices which address particular technical points discussed in the main text. 

\section{\label{sec:hamsym} The Hamiltonian and its symmetry}

The local Hilbert space of our model $\mathbb{C}^N$ is spanned by the $N$ components of the SU(N) group, or $N$ species of hard-core bosons.  The model is defined on a one-dimensional lattice of length $L$, with periodic boundary condition $\mathbb{C}_{L+1}\equiv\mathbb{C}_1$. Its Hamiltonian takes the form 
\begin{equation}
\label{eq:ham}
H=\sum_{i=1}^L \left[P_{i,i+1}+2\Delta C_{i,i+1}\right]
\end{equation}
where transposition operator $P_{i,i+1}$ and the diagonal anisotropy term $C_{i,i+1}$ are defined as 
\begin{align}
P_{i,i+1}=&\sum_{a,b=1}^N e_i^{(ab)}\otimes e_{i+1}^{(ba)},\\
C_{i,i+1}=&\sum_{a=1}^N e_i^{(aa)}\otimes e_{i+1}^{(aa)},
\end{align}
where $(e^{(ab)})_{cd}=\delta_c^a\delta_d^b$ is the standard basis of $N\times N$ matrices, such that $P_{i,i+1}(v_i\otimes v_{i+1})=v_{i+1}\otimes v_i$ for any $v_i,v_{i+1}$. The transposition term is SU(N) invariant, as it can be written as traces over product of SU(N) generators. The diagonal term however, can be written only in terms of the generators of the Cartan subalgebra of SU(N), and therefore breaks the symmetry down to the symmetric group $\frak{S}_N$. 


Apart from the multi-component chain interpretation, one can also look at this model as a chain of hard-core bosons with a local color degree of freedom subjected to a repulsive density-density interaction. Henceforward, we shall interchange freely between these two languages for the convenience of presentation. 
The formulation in terms of hard-core bosons permits to put in correspondence the present model with the number-theory quantum model analyzed in \cite{mussardo2020prime} (the so-called coprime spin ladder model), in the sense that each boson with different color can be put in correspondence with a prime number while the diagonal term in the Hamiltonian can be associated to the "coprime interaction". There are though some differences between the two models: unlike the one discussed in \cite{mussardo2020prime}, we do not have here a quasi-2D lattice and, moreover, in the present case each boson does not have a composite structure consisting of more elementary degrees of freedom. Moreover, there is no onsite dynamical terms in the Hamiltonian that allows bosons of one species to transmute into others. Nonetheless, the coprime spin ladder model does reproduce the Hamiltonian (\ref{eq:ham}) when restricted to a certain sector of particle content $\{c_1^{n_1}, c_2^{n_2},\cdots, c_s^{n_s}\}$, where $c_A$ denotes the particle of color $A$, and $n_A$ denotes the number of times it appear along the chain. Therefore, our Hilbert space is fragmented into Krylov sub-spaces specified by the particle content. In each sector, the restricted Hamiltonian will be equivalent to \cite{Sutherlandprb}
\begin{equation}
H(\{c_1^{n_1}, c_2^{n_2},\cdots, c_s^{n_s}\})=\sum_{i=1}^L\left[(1+2\Delta)\sum_A N_{i,i+1}^{(AA)}+\sum_{A<B} P_{i,i+1}^{(AB)}\right],
\label{eq:hamsec}
\end{equation}
where $N^{(AA)}$ counts the number of neighboring pairs of species $A$, and $P^{(AB)}$ transposes only neighboring pairs of species $A$ and $B$. 

This restricted Hamiltonian manifests permutation symmetries in addition to the translational invariance. Firstly, the total number of each species $n_A$ is conserved. Secondly, being each particle indistinguishable, the  Hamiltonian is invariant under $\frak{S}_{n_1}\otimes \frak{S}_{n_2}\otimes\cdots\otimes \frak{S}_{n_s}$, where $\sum_{A=1}^s n_A=L$, and $\otimes$ denote the outer product between symmetric groups \cite{hamermesh, ElliotDawber}. Notice that this is a subgroup of the original $\frak{S}_L$ symmetry. Therefore, eigenvectors of this Hamiltonian can be identified  within each of its irreducible representations (irrep's).  
This step would already reduce the complexity of the problem significantly enough so that one could diagonalize the Hamiltonian written in terms of standard irreducible representations matrices by brute force, as was done in \cite{PhysRevLett.113.127204,PhysRevB.93.155134}. With some more thoughts, Gaudin \cite{Gaudin:1967aa, Gaudinbook} turned the Young tableaux formalism (equivalently described in terms of Hund's method, usually more appealing to physicist), into a linear system of constraints from Fock's conditions, which by one remarkable algebraic identity after another, arrived at the same Bethe-Yang Hypothesis states. However, the anisotropy term in our Hamiltonian mixes irreducible representations, making this approach not applicable. So, in our case it is more promising to follow the strategy of Yang and Sutherland~\cite{yangprl,Sutherlandprl,Sutherlandprb} and first take advantage of the translational invariance to go to momentum space and consider irrep's of permutation operators.

\section{\label{sec:diag} Diagonalization of the integrable subspaces}

The Hamiltonian \eqref{eq:ham} without the anisotropy term was  diagonalized by Sutherland using nested Bethe Ansatz \cite{Sutherlandprb} \footnote{Although Sutherland was actually solving the ferromagnetic model for fermionic particles, he showed the spectrum is that of the antiferromagnetic bosonic one in reversed ordering.}. As shown in the next section, the anisotropy term breaks the integrability of the original Hamiltonin by violating the Yang-Baxter Equation (YBE) of scattering matrices, except for certain irrep's. The large number of local degrees of freedom, or species of particles, can be both a curse and a blessing. In this paper, we take advantage of this property by constructing a particular basis, in which the solution of relatively low-lying eigenstates using the Bethe Ansatz is drastically simplified. In this section, we first show how this approach helps solving eigenstates in the sector where only one specie of particle appear multiple times, while the rest of the species each appears only once. In the next section, we show that the YBE is still violated in this basis for a generic sector where multiple species appear more than once, as well as in the orthogonal subspace to the basis within the same symmetry sector.

\subsection{\label{sec:sing} Maximally antisymmetrized basis}

\begin{figure}
	\includegraphics[width=0.5\linewidth]{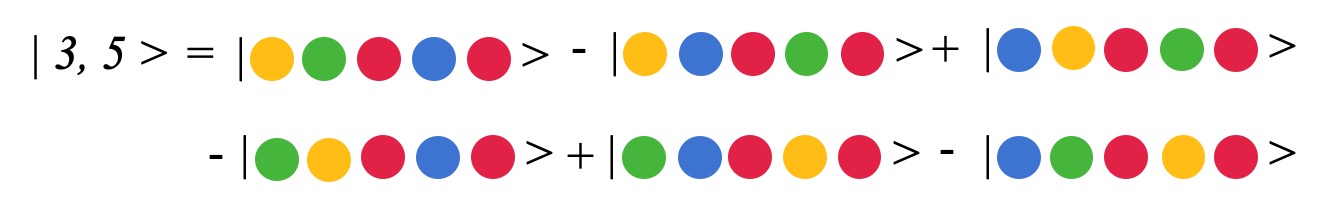}
	\caption{An example of the basis vectors in a sector with particles of $4$ different colors, the red onappearing twice.}
	\label{fig:basis}
\end{figure}

The key observation which leads to use this method consists of  realizing that each transposition operator is minimized by the eigenvalue $-1$. So the ground state corresponds to that state reached by antisymmetrizing as many neighbors as possible. Of course this can only reconcile among different species, as antisymmetrizing a symmetric pair gives $0$. Therefore, we can restrict our searching for the ground state in the subspace of the Hilbert space spanned by the basis
\begin{equation}
\label{eq:basis}
|{i_1,i_2,...,i_n}\rangle = \sum_{\sigma \in \frak{S}_{L-n}}\text{sgn}(\sigma) (-)^{\sum_{a=1}^n i_a} |{i_1,i_2,...,i_n};\sigma\left(c_2~c_3~\cdots~c_{L-n}\right)\rangle,
\end{equation}
where $i_1<i_2<\cdots<i_n$ labels the location of the $n$ identical species (using the correspondence with the coprime spin ladder model, these species are chosen to be the one corresponding to the smallest prime numbers). The symbol sgn($\sigma$) denotes the signature of the permutation $\sigma$ acting on the reference state in a certain order of species with $c_j$ denoting particle of the $j$-th color, before permuting the identical species to their final locations. An illustration of the $n=2, L=5$ case is given in Fig.~\ref{fig:basis}.
The reason for this convenient choice of basis is such that
\begin{equation}
\label{eq:P}
P_{j,j+1} |{i_1,...,i_n}\rangle =  \begin{cases} -|{i_1,...,i_n}\rangle, &j\ne i_a,i_a-1,\forall a\\  -|{i_1,...,i_{a-1},i_{a}+1,i_{a+1},...,i_n}\rangle,  &j=i_a\ne i_{a+1}-1 \\
|{i_1,...,i_n}\rangle, &j=i_a=i_{a+1}-1
\end{cases}
\end{equation}
for any $i_a<L$. Since the transposition operators act the kinetic term of the Hamiltonian, antisymmetrizing the species is equivalent to picking a subspace where all the particles with different species except the identical one that appears multiple times to have momenta $\pi$.
Periodic boundary condition on the physical chain leads to (anti-)periodic boundary condition on our artificial basis, depending on the parity of the length of the chain.

\begin{equation}
\label{eq:apbc}
|i_1,...,i_{n-1},L+1\rangle \equiv (-1)^L |1,i_1,...,i_{n-1}\rangle,
\end{equation}
where $i_1>1$. For simplicity, we restrict our discussion to the case of even length from now on. The odd length case can be treated likewise. 
In this basis, we can diagonalize the Hamiltonian by expanding the eigenvector as 
\begin{equation}
\label{eq:eigstate}
|v\rangle = \sum_{1\le i_1<\cdots<i_n\le L}w_{i_1,...,i_n} |{i_1,i_2,...,i_n}\rangle.
\end{equation}

Keeping in mind that our basis vectors are already eigenvectors of transposition operators unless it acts on the identical species, the eigenvalue equation 
\begin{equation}
\label{eq:eig}
H|v\rangle = E_v |v\rangle.
\end{equation}
results in a difference equation version of the Helmholtz equation of weights $w_{i_1,...,i_n}$
\begin{equation}
\label{eq:reccur}
\nabla^2 w_{i_1,...,i_n} = -(E_v+L) w_{i_1,...,i_n},
\end{equation}
where we used $\nabla^2$ to denote the discrete Laplacian operator so as to avoid confusion with the anisotropy parameter. If none of the position $i$'s is consecutive, 
\begin{equation}
\nabla^2 w_{i_1,...,i_n} \equiv\sum_{a=1}^n \left(w_{i_1,...,i_a-1,...,i_n}+w_{i_1,...,i_a+1,...,i_n}-2w_{i_1,...,i_n}\right),
\label{eq:nonadjhm}
\end{equation}
where $w_{i_1,...,i_{n-1},L+1}\equiv (-1)^{n-1}w_{1,i_1,...,i_{n-1}}$ due to \eqref{eq:apbc}. When certain position $i$'s form consecutive strings $l_a,l_a+1,...,r_a$, $\nabla^2$ becomes instead 
\begin{equation}
\label{eq:adjdelt}
\begin{split}
\nabla^2 w_{l_1,...,r_1,...,l_s,...,r_s}=\sum_{\alpha=1}^{s}\bigl( w_{l_1,...,r_1,...,l_\alpha-1,l_\alpha+1,...,r_\alpha,...,l_s,...,r_s} \\ +w_{l_1,...,r_1,...,l_\alpha,...,r_\alpha-1,r_\alpha+1,...,l_s,...,r_s}\\ - [\sum_{a=l_\alpha}^{r_\alpha-1}(2+2\Delta) +2]w_{l_1,...,r_1,...,l_s,...,r_s} \bigr),
\end{split}    
\end{equation}

Let's now make few remarks to underline the distinction which exists between our model and the ferromagnetic spin-1/2 Heisenberg model: in the Bethe Ansatz solution of this latter model, the coefficient of the third term in eq.~\eqref{eq:adjdelt} would remain $-2$, regardless of how long the string of consecutive spin downs would be. This is the main difference between the two models which will eventually lead to our different solution of the Bethe Ansatz equations, as explicitly illustrated below. One might think that our derivation of Bethe Ansatz equations, obtained from subtracting these two equations and  requiring them both to hold simultaneously, would put a restriction to our solution, namely  that the number of components must exceed half of the size of the system for \eqref{eq:nonadjhm} to be valid. However, a closer examination reveals that the subtraction is merely a shortcut in deriving the Bethe Ansatz equations, since  alternatively one can compare the coefficients of independent monomials on both sides of \eqref{eq:adjdelt} and get the same equations. That is to say, our solution exists for arbitrary number of components $N$.

\subsection{Elementary two-body example: the $\{c_1^2,c_2,\cdots,c_{L-2}\}$ sector}
\label{sec:2bdy}
\subsubsection{The isotropic case}
The following two subsections serve as a very elementary introduction to readers who may be not familiar with the Bethe Ansatz formalism. More experienced readers should jump directly to our main result in \ref{sec:mb}. The case of $n=2$ as we will see shortly corresponds to a 2-quasi-particle excitation from the background of momentum-$\pi$ species. Considering this case is particularly illuminating as it allows an exact diagonalization by solving a second-order linear homogeneous recurrence relation. To see this, we perform the transformation
\begin{equation}
w_{i_1,i_2}\rightarrow a_r^{(s)}=w_{r,r+s},
\end{equation}
where $r$ labels the site of the first occurrence of the identical species, while $s$ denotes the distance between the two of them.
The eigenvector is then expressed as
\begin{equation} |v\rangle = \sum_{r=1}^L\sum_{s=1}^{L-r} a_r^{(s)} |{r, r+s}\rangle. \end{equation}
And the anti-periodic boundary condition is given by 
\begin{equation}\label{eq:2apbc} a_r^{(L-r+1)}=-a_1^{(r-1)}.\end{equation}

We first exploit the translational invariance of the Hamiltonian by diagonalizing in the basis formed by eigenvectors of the translation operator
\begin{equation}
\label{eq:trans}
T|v\rangle = t^{-1}|v\rangle.
\end{equation}
According to the anti-periodic boundary condition \eqref{eq:apbc}, we have 
\begin{equation}
T^L|1,2\rangle=T^2|L-1,L\rangle=-T|1,L\rangle=|1,2\rangle.
\end{equation} So $T^L=1$ still applies, giving 
\begin{equation}
\label{eq:mom}
t=e^{2i\theta},\quad \theta=\frac{i \pi k}{L}, \quad k=1,...,L.
\end{equation}
Therefore, \eqref{eq:trans} implies $a_{r+1}^{(s)}=t a_r^{(s)}$, and up to a normalization constant, we can choose $a_r^{(1)}=t^r$ as the initial condition for recurrence equations of $a_r^{(s)}$ with respect to index $s$.

The recurrence relation \eqref{eq:reccur} now becomes
\begin{align}
\label{eq:ini2}
a_r^{(2)}+a_{r-1}^{(2)} =& e a_r^{(1)} \\ 
a_{r+1}^{(s-1)}+a_r^{(s-1)}+a_r^{(s+1)}+a_{r-1}^{(s+1)}=& e a_r^{(s)}, \quad s>1,
\end{align}
where $e=-(E_v+N-4)$. Using translation invariance, the first equation gives the second initial condition necessary to solve the second-order recurrence relation $a_r^{(2)}= \frac{e t^{r+1}}{1+t}$, while the second equation becomes 
\begin{equation}
\label{eq:2reccur}
a_r^{(s+1)}-\frac{et}{1+t} a_r^{(s)} + t a_r^{(s-1)}=0.
\end{equation}

The solution of its characteristic equation depends on its discriminant
\begin{equation}
D=\frac{e^2t^2}{(1+t)^2}-4t \,\,\,.
\end{equation} 
\paragraph{$D=0$: Identical root:} 
In this scenario, the general solution is of the form $a_r^s=(c_0 +c_1 s)t^s$, which cannot satisfy the anti-periodic boundary condition \eqref{eq:2apbc} and the initial conditions simultaneously. We conclude that the discriminate must be nonzero.

\paragraph{$D\ne0$: Different root:} 
The characteristic equation has two roots in this case
\begin{align}
\lambda_{\pm}=&\left(\frac{e}{4\cos\theta}\pm i \sqrt{1-\left(\frac{e}{4\cos\theta}\right)^2}\right)e^{i\theta}\\
=&e^{i(\theta\pm\alpha)},
\end{align}
where $\alpha=\arccos(\frac{e}{4\cos\theta})$. Here, we have assumed $|e|\le4|\cos\theta|$, which is necessary for the anti-periodic boundary to hold.
Together with the initial conditions given by $a_r^{(1)}$ and $a_r^{(2)}$, \eqref{eq:2reccur} gives
\begin{equation}
\label{eq:2coeff}
a_r^{(s)}=t^{r+(s-1)/2}\frac{\sin s\alpha}{\sin\alpha}.
\end{equation}
Plugging this into the anti-periodic boundary condition \eqref{eq:2apbc}, we get
\begin{equation}
\alpha = \theta + \theta',\quad \theta'= \frac{2\pi k'}{L},\quad k'=1,2,...,L.
\end{equation}
We can parameterize the energy eigenvalues with $\theta_1\equiv \theta -\alpha=-\theta'$ and $\theta_2\equiv \theta+\alpha =2\theta+\theta'$, and the eigenvectors with $\mu_{1,2}\equiv e^{i\theta_{1,2}}$ as
\begin{align}
\label{eq:gsenergy}
E(\theta_1,\theta_2)=&-L+4-2\cos \theta_1-2\cos\theta_2\\ 
\label{eq:gsstate}
|\theta_1,\theta_2\rangle=&\frac{1}{N}\sum_{1\le i<j\le L}(\mu_1^i\mu_2^j-\mu_1^j\mu_2^i)|i,j\rangle,
\end{align}
where $\theta_{1,2}=i 2\pi k_{1,2}/N$, with $k_{1,2}=1,2,...,N$. Since its coefficients $a_{i,j}=-a_{j,i}$, we can alternatively express it as 
\begin{equation}
|\theta_1,\theta_2\rangle=\frac{1}{N}\sum_{1\le i,j\le L}\mu_1^i\mu_2^j|i,j\rangle,
\end{equation}
with $|i,i\rangle\equiv0$. Notice the $\theta_1=\theta_2$ solutions give vanishing eigenvectors, as is expected when one attempts to antisymmetrize a symmetrized pair of identical species. In other words, when the anisotropy or density-density interaction is absent, these identical particles of same species behave as free fermions. The fact that our original bosonic degrees of freedom has ended up behaving as fermions is exactly due to the unitary transformation pointed out by Sutherland \cite{Sutherlandprb} of multiplying all wave functions by completely anti-symmetric one in all objects, which flips the sign of the Hamiltonian and the exchange statistics of the particles at the same time.
The ground state energy in this sector is therefore $-L+2-2\cos\frac{2\pi}{L}$. In the thermodynamic limit, it gives rise to a vanishing gap that scales with $1/L^2$ counting from the universal ground state of $L$ different species.

The partial spectrum and corresponding eigenstates given above agree with the results from nested Bethe Ansatz by Sutherland \cite{Sutherlandprb}, which starts from a reference state of identical species and consider the different ones as quasi-particles moving around. To map the eigenstates solved from our approach of treating the different species as background and identical ones as quasi-particles, to the framework of Bethe Ansatz, the different species are understood to have $\pi$ momenta. Our solution corresponds to two particles having momenta $\theta_1$ and $\theta_2$, with a scattering phase of $\pi$, which says the two identical species albeit impossible to be antisymmetrized in real space, is antisymmetrized in the momentum space.

\subsubsection{Two-body problem with anisotropy/interaction}
Since the interaction term acts on neighboring sites, it would only change the $e$ in \eqref{eq:ini2} to $e+2\Delta$ in determining the initial condition $a_r^2$, leaving the recurrence equation unchanged. Their solution is given by 
\begin{align*}
a_r^{(s)}(\Delta)=&\frac{t^{r+(s-1)/2}}{2i\sin\alpha}\left[\left(\frac{e+2\Delta}{2\cos\theta}-e^{-i\alpha}\right)e^{i(s-1)\alpha}+\left(e^{i\alpha}-\frac{e+2\Delta}{2\cos\theta}\right)e^{-i(s-1)\alpha}\right]\\
=& t^{r+(s-1)/2}\left(\frac{\sin(s\alpha)}{\sin\alpha}+\frac{\Delta}{\cos\theta}\frac{\sin(s-1)\alpha}{\sin\alpha}\right),
\end{align*}
which reproduces \eqref{eq:2coeff} when $\Delta$ is taken to be $0$.
Quantization of the energy is once again given by the solution of the anti-periodic boundary condition, now written as
\begin{equation}
e^{i L(\alpha-\theta)}=\frac{\cos\theta+\Delta e^{i\alpha}}{\cos\theta+\Delta e^{-i\alpha}}
\end{equation}
Taking the logarithm on both sides, one realize that for each value of $\theta=\frac{k\pi}{N}$, there are $N$ roots $\alpha_J$, labeling the $N$ degenerate eigenstates at energy level $E(k,J)=-N+4-4\cos \theta \cos \alpha_J$,
\begin{equation}
L(\alpha_J-\theta)+2\pi J=2\arctan\frac{\Delta\sin\alpha_J}{\cos\theta+\Delta\cos\alpha_J}, \quad J=1,2,...,L.
\end{equation}
Parametrizing as before, $\theta_1\equiv \theta -\alpha$, $\theta_2\equiv \theta+\alpha$, and $\mu_{1,2}\equiv e^{i\theta_{1,2}}$, we still have
\begin{align}
E(\theta_1,\theta_2)=&-L+4-2\cos \theta_1-2\cos\theta_2\\ 
|\theta_1,\theta_2\rangle=&\frac{1}{\mathcal{N}}\sum_{1\le i<j\le L}(\mu_1^i\mu_2^j-S_{12}\mu_1^j\mu_2^i)|i,j\rangle,
\end{align}
where $S_{12}=\mu_1^{-L}\equiv \mu_2^L$, since $(\mu_1\mu_2)^L\equiv1$, and normalization constant $\mathcal{N}^2=L^2-L-L e^{i L\theta} \sin (L-1)\alpha/ \sin \alpha$.

\subsection{\label{sec:manybody} The multi-mode scars: $\{c_1^{n},c_2,\cdots,c_{L-n+1}\}$ sector}
\label{sec:mb}

When there are more than two identical particles in one species, the recursive equations as in the previous section will involve multiple indices. Therefore, it is necessary to construct the Bethe Ansatz wave function 
\begin{equation}
\label{eq:coeffba}
w_{i_1,...,i_n}=\sum_{\sigma\in \frak{S}_n}A_\sigma \prod_{a=1}^n \mu_{\sigma a}^{i_a},
\end{equation}
where $\frak{S}_n$ denotes symmetric group of order $n$, and $\mu_a=e^{i\theta_a}$.
When $i_{a+1}>i_a+1$ for all $a$, \eqref{eq:reccur} takes the form
\begin{equation}
\label{eq:nonadj}
\sum_{\sigma\in \frak{S}_n} A_\sigma \sum_{a=1}^n (\mu_{\sigma a}^{-1} + \mu_{\sigma a}-2)\prod_{a=1}^n \mu_{\sigma a}^{i_a}=-(E_v+L)\sum_{\rho\in S_n} A_\rho \prod_{a=1}^n \mu_{\sigma a}^{i_a}.
\end{equation}
The common factor on the l.h.s. can be taken out of the sum over $\sigma$, which gives $E_v=-L+\sum_{a=1}^n(2-2\cos \theta_a)$.
When $i_{a_0+1}= i_{a_0}+1$ for 
some $a_0$, but $i_{a+1}>i_{a}+1$ for $a\ne a_0$,\eqref{eq:reccur} becomes
\begin{equation}
\label{eq:adj}
\begin{split}
\sum_{\sigma\in \frak{S}_n} A_\sigma \bigl(\mu_{\sigma a_0}^{-1} + \mu_{\sigma (a_0+1)}-4-2\Delta+\sum_{a\ne a_0}(2\cos \theta_{\sigma a}-2)\bigr)\prod_{a=1}^n \mu_{\sigma a}^{i_a} \\    =-(E_v+L)\sum_{\rho\in S_n} A_\rho \prod_{a=1}^n \mu_{\rho a}^{i_a}.
\end{split}
\end{equation}
Noticing that \eqref{eq:nonadj} still holds in this case, we can subtract this equation from it to get
\begin{equation}
\sum_{\sigma\in \frak{S}_n} A_\sigma (\mu_{\sigma a_0} +\mu_{\sigma(a_0+1)}^{-1} +2\Delta)\prod_{a=1}^n \mu_{\sigma a}^{i_a}=0,
\end{equation}
or using explicitly $i_{a_0+1}=i_{a_0}+1$,
\begin{equation}
\sum_{\sigma\in \frak{S}_n} A_\sigma \bigl(\mu_{\sigma a_0}\mu_{\sigma (a_0+1)} +1 + 2\Delta \mu_{\sigma(a_0+1)}\bigr)\mu_{\sigma (a_0+1)}^{-1}\prod_{a=1}^n \mu_{\sigma a}^{i_a}=0.
\end{equation}
Combining terms from permutations differing by a transposition $(a_0\ a_0+1)$, one can see that for this equation to be satisfied for any coordinates $i_a$'s and momenta $k_a$'s, it requires
\begin{equation}
\label{eq:scatterph}
\frac{A_{\sigma'}(\Delta)}{A_\sigma(\Delta)}=-\frac{ \mu_{\sigma a_0}\mu_{\sigma(a_0+1)} +1+2\Delta\mu_{\sigma (a_0+1)}}{\mu_{\sigma a_0}\mu_{\sigma(a_0+1)} +1+2\Delta\mu_{\sigma a_0}},
\end{equation}
where $\sigma'=\tau_{a_0}\sigma$, with $\tau_{a_0}$ transposing $a_0$ and $a_0+1$. In appendix \ref{sec:parametrization}, we introduce different parameterization of the momenta for different magnitudes of the anisotropy, in which the scattering phase depends only on the difference of the rapidities of the two scattering particles, so as to facilitate a unified treatment of nested and algebraic Bethe Ansatz in the next section.
Quantization condition comes from the periodic boundary condition \eqref{eq:apbc}. Decomposing the cycle-$n$ permutation $\tau$ into a product of transpositions between neighbors, and using \eqref{eq:scatterph}, we have
\begin{equation}
\prod_{a=1}^{n} \frac{\Delta\cos\frac{1}{2}(\theta_{\sigma n}-\theta_{\sigma a})+\cos\frac{1}{2}(\theta_{\sigma n}+\theta_{\sigma a})+i\Delta\sin\frac{1}{2}(\theta_{\sigma n}-\theta_{\sigma a})}{\Delta\cos\frac{1}{2}(\theta_{\sigma n}-\theta_{\sigma a})+\cos\frac{1}{2}(\theta_{\sigma n}+\theta_{\sigma a})-i\Delta\sin\frac{1}{2}(\theta_{\sigma n}-\theta_{\sigma a})} = e^{i\theta_{\sigma n}L},
\end{equation}
where we have absorbed the minus signs into the scattering phases. Since \eqref{eq:quantization} holds for any $\sigma \in S_n$, we have the Bethe Ansatz equations
\begin{equation}
2\sum_{b=1}^n \arctan \frac{\Delta\sin\frac{1}{2}(\theta_a-\theta_b)}{\Delta\cos\frac{1}{2}(\theta_a-\theta_b)+\cos\frac{1}{2}(\theta_a+\theta_b)}=L\theta_a+2\pi J_a
\end{equation}
where $J_a=1,...,L$, for any $a=1,...,n$.
Taking the sum on both sides over $a$, we get $\sum_{a=1}^n \theta_a=2\pi k/L$, with $k=1,...,L$.

Once the momenta are solved from the Bethe Ansatz equations, \eqref{eq:scatterph} can be used to deduce the relative weights of the permutations among momenta, which gives the eigenstates in this sector.
If the interaction term is turned off, \eqref{eq:scatterph} becomes
\begin{equation}
A_\sigma=-A_{\sigma'}.
\end{equation}
Therefore $A_\sigma=(-1)^\sigma$. Boundary condition now becomes
\begin{equation}
\label{eq:quantization}
\sum_\sigma A_\sigma \mu_{\sigma n}^{L+1}\prod_{a=1}^{n-1} \mu_{\sigma a}^{i_a} =(-1)^{n-1}\sum_\rho A_\rho \mu_{\rho 1} \prod_{a=2}^{n} \mu_{\rho a}^{i_a}, 
\end{equation}
for any $i_1,...,i_n$. Comparing the terms on both sides differing by a global translation $\tau=(1\ 2\ \cdots\ n)$, we have
\begin{equation}
\label{eq:freebc}
(-1)^{n-1}=\frac{A_{\sigma \tau^{-1}}}{A_\sigma}=(-1)^{n-1}\mu_{\sigma n}^{L}, \quad \forall \sigma\in S_n.
\end{equation}
Hence, $\theta_a=2\pi k_a/L$, and the ground state energy in the sector with $n$ particles belonging to the identical species is 
\begin{equation} 
E\, =\,-L + 2 \sum_{k=-m}^m\left(1 -\cos 2\pi k/L\right)
\end{equation}
for $n=2m+1$, or 
\begin{equation}
    E = -L + 2 \sum_{k=-m+1}^m\left(1-\cos 2\pi k/L\right)
\end{equation}
for $n=2m$, since if any two of the momenta coincide, the anti-symmetrized coefficient \eqref{eq:coeffba} would vanish. We can use the Lagrange's trigonometric identity to estimate their gap from the ground state in the thermodynamic limit $n\to\infty$, \begin{equation} E-E_0 \to 2L \left(\frac{n}{L}-\frac{2}{\pi}\sin \frac{\pi n}{L}\right)\end{equation} 
With some extra steps explained in appendix \ref{sec:norm}, one can even show that the normalization constant is fixed to be $\mathcal{N}=L^{n/2}$. The eigenstates are then expressed as
\begin{equation}
|\theta_1,...,\theta_n\rangle=L^{-n/2}\sum_{1\le i_1<\cdots<i_n\le L}\sum_{\sigma\in \frak{S}_n}(-1)^\sigma e^{i\sum_{a=1}^n \theta_{\sigma a}i_a}|i_1,...,i_n\rangle.
\end{equation}
The careful reader might now recognize that, choosing $n$ momenta among their $L$ possible values allowed by quantization condition, the dimension of this eigenvector subspace $\binom{L}{n}$, is precisely the dimension of the irreducible representation corresponding to the one-column Young tableaux with $n$ rows, whereas the absolute ground state of the full Hilbert space expressed by Slater determinant correspond to the $1$ dimensional irreducible representation of the $L$ row, $1$ column tableaux.

We emphasize that the integrable eigenstates found above have finite energy density (evaluated wrt the ground state energy) in the thermodynamic limit. To see that, notice that the energy of eigenstates in this sector consists of two parts, a contribution from the antisymmetrized basis, at the lowest energy $\epsilon_{\text{GS}}=-L$, and a contribution from the solution of spin-$\frac{1}{2}$ XXZ problem, of the form $L\epsilon_{\text{XXZ}}(S^z_{\text{tot}}=\frac{L}{2}-n) $, where lower and upper bound on $\epsilon_{\text{XXZ}}(S^z_{\text{tot}}=\frac{L}{2}-n)$ is established to be finite~\cite{PhysRev.147.303}. So the total energy density in the thermodynamic limit is
\begin{equation}
	\epsilon_v-\epsilon_{\text{GS}}= \epsilon_{\text{XXZ}}.
\end{equation}

\section{\label{sec:multi} Violation of YBE: the $\{c_1^{n_1},\cdots,c_m^{n_m},c_{m+1},\cdots,c_{L-n+m}\}$ sector}

The spectrum of a generic sector, where multiple colors appear multiple times in the particle content, is solved by recourse to the Bethe-Yang hypothesis \cite{yangprl}, used in Yang's solution to the one-dimensional Fermi problem with repulsive $\delta$ interaction. As well known,  the same model was originally and independently solved by Gaudin, using a less physical and more algebraically involved approach \cite{Gaudin:1967aa},) based on the pioneering work of McGuire \cite{McGuire:1965aa,McGuire:1966aa}, Lieb and Flicker \cite{PhysRev.161.179}, and subsequently further generalized by Sutherland \cite{Sutherlandprl, Sutherlandprb}.   

Yang's approach differs from Bethe Ansatz used in the previous section since it does not assume any constraint on the permutation symmetry of the wave function, which is indeed the case when multiple species are involved. Therefore, if we still label the sites of all identical species with $i$'s, eq.\, \eqref{eq:P} no longer holds. Consequently, the basis $|i_1,\cdots,i_n\rangle$ no longer spans the whole Hilbert space when restricted to $i_1<\cdots<i_n$. Instead, we should write the eigenvector of the Hamiltonian as 
\begin{equation}
	\label{eq:multieig}
	|v\rangle=\sum_{Q\in \frak{S}_n}\sum_{1\le x_{Q1}<\cdots,x_{Qn}\le L} \psi_Q(\mathbf{x})|\mathbf{x}\rangle,
\end{equation}
where $\mathbf{x}$ is an $n$-component array of the combinations of coordinate and spin/color $\{q_i;c_i\}$, and $x_i<x_j$ if $q_i<q_j$, or $c_i<c_j$ if $q_i=q_j$. The simultaneous swapping of a pair both of them gives a minus sign due to the fermionic nature of the basis,
\begin{equation}
	\label{eq:multisgn}
	|\mathbf{x}\tau\rangle=\text{sgn}(\tau)|\mathbf{x}\rangle,
\end{equation}
where $\mathbf{x}\tau=(x_{\tau(1)},...,x_{\tau(n)})\equiv(\{q_{\tau(1)};c_{\tau(1)}\},...,\{q_{\tau(n)};c_{\tau(n)}\})$. Therefore, only the anti-symmetric part of the wave function multiplying an anti-symmetric basis is meaningful
\begin{equation}
	\label{eq:multiwf}
	\psi_{\tau^{-1} Q}(\mathbf{x}\tau)=\text{sgn}(\tau)\psi_Q(\mathbf{x}).
\end{equation}

While we are discussing permutation symmetry, it's worth mentioning theorems on the ordering of energy levels of the isotropic Hamiltonian. According to Lieb and Mattis' theorem on the ordering of energy levels of antiferromagnets, the lowest energy eigenstate is determined by the so-called ``pouring principle" \cite{LiebMattiasferro, Liebhubbard}. More recently, this result has been  generalized to a lattice model with higher spin with both open and periodic boundary conditions \cite{HAKOBYAN2004575}. In short, it states that the lowest energy eigenstate corresponds to the Young Tableaux that gives the highest weight state of the $sl(n)$ algebra. The proof is given much in the same spirit of the one due to Lieb and Mattis, namely, showing that the representation corresponding to the Young Tableaux of highest weight state follows from the Perron-Frobenius theorem and the requirement to have non-negative components: hence, it overlaps with the ground state, but, given at the same time that the Hamiltonian does not mix representations, one can conclude  that it has to be the only representation corresponding to the ground state. 
Irreducible representations of outer product can be constructed according to a simple rule from the irreducible representations of each group in the outer product, which in our case, has to be the single row Young tableaux $[n_1],\cdots,[n_m]$, as one cannot antisymmetrize among an identical species. Then the Young Tableaux in the outcome of the outer product that every other one can ``pour into'', alias corresponding to the highest weight state, is $[n_1,\cdots,n_m]$, where each row is filled with the same species, assuming $n_1>n_2>\cdots>n_m$. 
The above analysis relies purely on the permutation symmetry of the Hamiltonian. As we will see shortly, Yang's approach utilizes both permutation symmetry and the translational invariance, and is more convenient in practice.
It starts with a first layer of Bethe Ansatz treating all particles as if they have the same color, while keeping in mind that their colors could be different later by keeping track of the swapping of color indices
\begin{equation}
	\psi_Q(\mathbf{x}) =\sum_{P\in \frak{S}_n} A_{Q,P} \prod_{i=1}^n \mu_{Pi}^{q_{Qi}}, 
\end{equation}
for each sector $1\le x_{Q1}<x_{Q2}<\cdots<x_{Qn}\le L$ in the coordinate space labeled by a permutation $Q\in \frak{S}_n$, which are a priori independent. Each $Q$ corresponds to a particular sector of the whole coordinate space, with the $Q=id$ one called the fundamental sector. $A_{Q,P}$ is an $n!\times n!$ matrix, whose columns are denoted by $\xi_P$. To relate the wave functions defined in these separate sectors, we use the boundary conditions between two sectors $Q'=Q\tau_{a}$ that differ by a transposition between indices $a$ and $a+1$. In appendix \ref{sec:Smat}, we give the detailed derivation of Yang's scattering matrix $Y$ from comparing the eigenvalue equations of nonadjacent and adjacent cases, which requires
\begin{equation}
	\label{eq:Ydef}
	\xi_{P\sigma_{ij}}=Y_{ij}^{a}\xi_P,
\end{equation}
for $P(a)=i,P(a+1)=j$, where the $Y$ matrix as defined in Appendix \ref{sec:Smat}. These $n!(n-1)$ equations are consistent for the isotropic Hamiltonian only. For completeness, we carry out its diagonalization following Sutherland \cite{Sutherlandprb}.

\begin{figure}
	\includegraphics[width=0.8\linewidth]{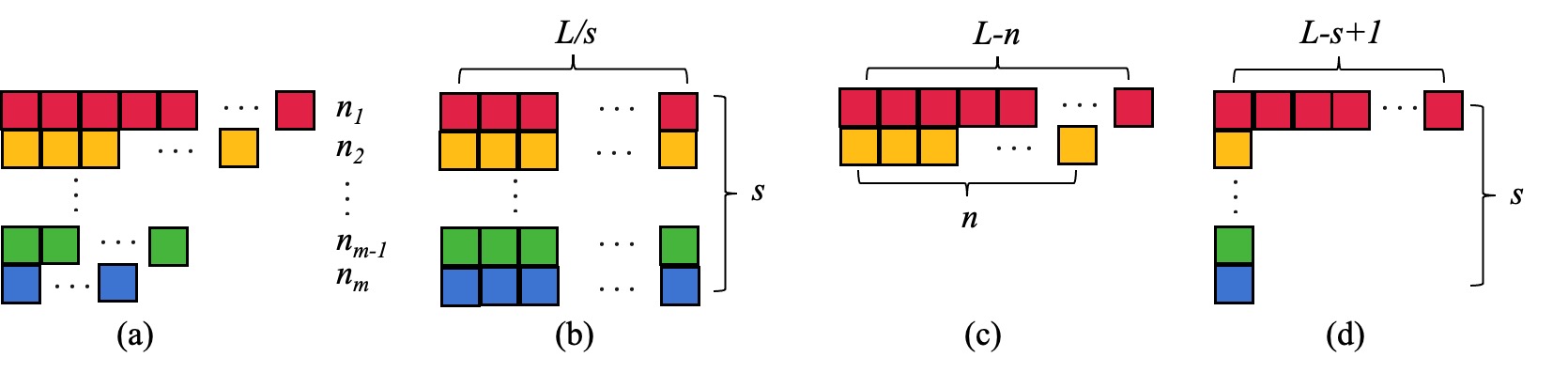}
	\caption{(a) Young diagram corresponding to a generic sector in the Hilbert space of our model, integrable only in the isotropic case. (b) The Young diagram corresponding to the sector where our ground state of a model with only a finite number $s$ of colors divisible by the size of the chain, not integrable as anisotropy mixes irreducible representations corresponding to different Young tableaux. (c) Young diagram corresponding to spin-$\frac{1}{2}$ XXZ model as well as the partially integrable subspaces studied in~\cite{Kiwata:1994vs,Sato:1995we,Sato:1996vm}. (d) Young diagram corresponding to the scarred sectors of our model.}
	\label{fig:Youngtab}
\end{figure}

The above Yang's $Y$ matrices are given in the \textit{reflection}(-diagonal) representation of the asymptotic wave function, according to Sutherland \cite{Sutherlandbook}, which is convenient for relating the a priori independent wave functions defined in separate sectors, as the coordinate of each particle remains invariant after scattering in this representation. However, in order to utilize the periodic boundary condition as a quantization condition for the momenta, we switch to the \textit{transmission}(-diagonal) representation, where the diagonal terms of scattering $S$  matrices are the transmission amplitudes. In the transmission representation, momentum is fixed to a particle after scattering. Picking a particular permutation of momenta $P_0=Q$, such that $i=a, j=a+1$, we have
\begin{equation}
	S_{ij}=\pi(\tau_{ij})Y_{ij}^{ij}=T_{ij}\mathbb{1}+R_{ij}\pi(\tau_{ij}),
\end{equation}
where reflection $R_{jj}$ and the transmission $T_{jj}$ coefficients are defined as in eqs.~\eqref{eq:transcoeff},\eqref{eq:reflcoeff} herefater. In Appendix~\ref{sec:Smat}, we show that in the presence of anisotropy in the Hamiltonian, the Yang-Baxter's equation, which is necessary for the following procedures to work, is only satisfied for sub-sectors of the Hilbert space corresponding to two types of Young tableaux, namely those of the shape illustrated in Fig.~\ref{fig:Youngtab} (c) and (d). The former corresponds to the solution of spin-$\frac{1}{2}$ XXZ model, as well as the partially integrable subspaces studied in~\cite{Kiwata:1994vs,Sato:1995we,Sato:1996vm}. The latter corresponds to our scarred sectors. In fact, the Young tabeaux in Fig.~\ref{fig:Youngtab} (c) correspond to an even broader class of integrable eigenstate: while the antisymmetric bases requires the constituents to be all distinguishable, a symmetric basis does not require its constituents to be identical. So we can also form symmetric bases from those different components and obtain the upper spectrum counterparts of the eigenstates solved in the previous section. The important point is that we can either form totally symmetric or antisymmetric irrep's, but not a mixture of them in order for the YBE to be satisfied.

\begin{figure}
	\includegraphics[width=0.5\linewidth]{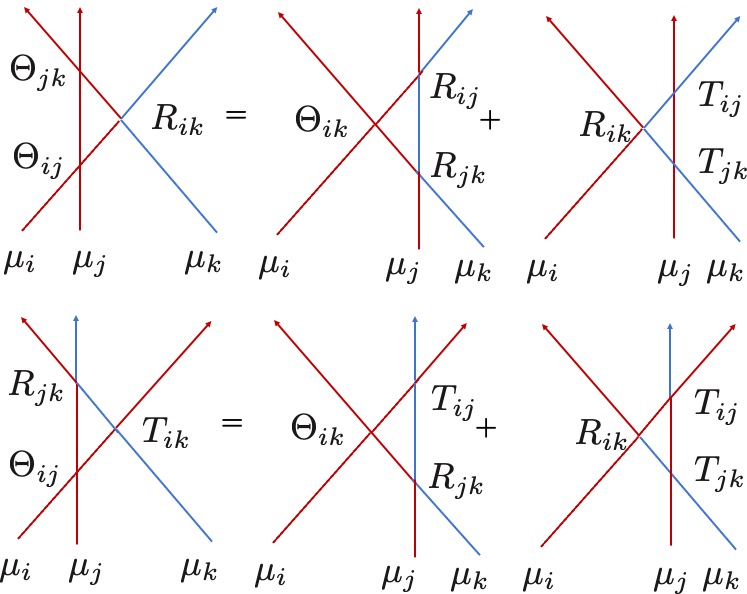}
	\caption{Yang-Baxter Equation in terms of reflection and transmission coefficient.}
	\label{fig:YBE}
\end{figure}

Hence, for scattering between different colors, $c_a\ne c_{a+1}$, we have
\begin{align}
	R_{ij}^{ab}(\Delta)&=\frac{-i}{\lambda_i-\lambda_j+i}\equiv R_{ij} \label{eq:reflcoeff}\\
	T_{ij}^{ab}(\Delta)&=\frac{\lambda_i-\lambda_j}{\lambda_i-\lambda_j+i}\equiv T_{ij} \label{eq:transcoeff},
\end{align}
where
\begin{equation}
	\lambda_i=\frac{i}{2}\frac{\mu_i+1}{\mu_i-1}.
\end{equation}
Yet for scattering between the same colors, $c_a=c_{a+1}$, we have $\pi(\tau_{a(a+1)})=-\mathbb{1}$ (due to \eqref{eq:multiwf}), and
\begin{equation}
	Y_{ij}^{a(a+1)}(\Delta)=R_{ij}^{a(a+1)}(\Delta)-T_{ij}^{a(a+1)}(\Delta)=-\frac{\mu_i\mu_j+1+2\Delta\mu_j}{\mu_i\mu_j+1+2\Delta\mu_i}=\frac{\varphi^{\Delta}(\lambda_j^\Delta -\lambda_i^\Delta+ i\eta)}{\varphi^{\Delta}(\lambda_j^\Delta-\lambda_i^\Delta - i\eta)}\equiv\Theta_{ij},
\end{equation}with the parameterization \eqref{eq:paramu}.
It's easily verified that while the unitarity relation 
\begin{equation}
	Y_{ij}^{ab}(\Delta)Y_{ji}^{ab}(\Delta)=1, 
\end{equation}
is satisfied for both cases, the Yang-Baxter's equations
\begin{equation}
	Y_{jk}^{ab}(\Delta)Y_{ik}^{bc}(\Delta)Y_{ij}^{ab}(\Delta)=Y_{ij}^{bc}(\Delta)Y_{ik}^{ab}(\Delta)Y_{jk}^{bc}(\Delta),
\end{equation} requires\begin{align}
	\Theta_{jk}R_{ik}\Theta_{ij}&=R_{ij}\Theta_{ik}R_{jk}+T_{ij}R_{ik}T_{jk},\\
	R_{jk}T_{ik}\Theta_{ij}&=T_{ij}\Theta_{ik}R_{jk}+R_{ij}R_{ik}T_{jk},
	\label{eq:YBE}
\end{align}which only hold when $\Delta=0$ or $\Delta\to\infty$, see Fig.\ref{fig:YBE}. 
If $\Delta\ne 0$, YBE hold respectively for scattering between identical colors, and different colors, but not when they are mixed. This means the scattering among three particles and more, is only factorizable into consecutive scatterings between pairs of two particles, irrespective of their ordering, if their colors are either all the same, which leads to the solution in \ref{sec:manybody}, or all different, corresponding to the absolute ground state. Hence, for the generic case, where scattering between particles of both same and different colors are present, the scattering is not non-diffractive, and Yang's nested Bethe Ansatz no longer applies. In fact, Sutherland has derived a necessary consistent condition for the scattering between different species to be non-diffractive  \cite{Sutherlandbook}, an example of such an exactly solvable anisotropic generalization of the multi-component Heisenberg model was solved by Babelon, et. al. \cite{Babelon}, albeit being less physically natural than our model.

Another way to look at the YBE is to find the momenta that solves Eq.~\eqref{eq:YBE} for generic $\Delta$. Multiplying the denominators on both sides and comparing the coefficients of different orders of $\Delta$ give the following solutions
\begin{equation}
\begin{cases}
	\mu_k = 1\\
	\mu_j=\mu_i;\\
	\mu_k=\mu_j;\\
	\mu_k=\mu_i;\\
	\mu_j=2-\frac{1}{\mu_i};\\
	\mu_k=2-\frac{1}{\mu_j};\\
	\mu_k=2-\frac{1}{\mu_i}.	
\end{cases}
\end{equation}
The first solution is trivial, meaning the particle with a different color has momentum $\pi$, or is antisymmetrized. This can only happen is appears only once along the chain for the wave-function not to vanish. So this is included in the case discussed before. The next three solutions corresponds to two of the three particles having the same momentum, in which case they move collectively keeping their distance, so a three-body scattering never happens. The last three solutions are non-trivial. They require two of the momenta to be complex-valued, with the corresponding particles forming bond states. 
\section{\label{sec:entropy}Entanglement entropy of integrable eigenstates}

 By now it should be clear that our integrable eigenstates are in 1-to-1 mapping with the Yang's solution to the $\delta$-interaction problem~\cite{yangprl} of $n$ particles, with the antisymmetrized bases of distinguishable particles playing the role of the vacuum there, or the pseudovaccum of all spin up state in the XXZ chain.
 In this section, we try to get a sense of the additional entanglement that these antisymmetrization contribute, by first show that in the extreme case of $N=L$, where the antisymmetrized basis occupy the entire system, the ground state has exactly linear scaling of entanglement entropy in the thermodynamic limit. Then we argue that the entanglement entropy of integrable excited states are lower-bounded by the corresponding Bethe Ansatz states in spin-$\frac{1}{2}$ XXZ chain~\cite{Alba_2009,Chen_2013}.

\subsection{Ground state entanglement entropy for $N=L$}

In the case of $N=L$, the ground state can be expressed as \begin{equation}
|GS\rangle = \frac{1}{\sqrt{L!}}\sum_{\mathcal{P} \in \frak{S}_L}\text{sgn}(\mathcal{P}) |c_{\mathcal{P}1}, c_{\mathcal{P}2}, ..., c_{\mathcal{P}L}\rangle,
\end{equation}
where $|..., c_i,...\rangle$ denotes the configuration with the $i$-th site having particle of color $c_i$. 

This ground state is highly entangled as the color at each site depends on those at all the others. This can be seen by calculation the entanglement entropy between two halves of the system. Let $L=2l$, then the Schmidt decomposition of the ground state is \begin{equation}
|GS\rangle=\sum_{\mathcal{P}_0 \in \frak{S}_L/(\frak{S}_l)^2} \text{sgn}(\mathcal{P}_0)
\frac{l!}{\sqrt{(2l)!}}|\mathcal{P}_0 1,\mathcal{P}_0 2,...,\mathcal{P}_0 l\rangle \otimes |\mathcal{P}_0(l+1),\mathcal{P}_0(l+2),...,\mathcal{P}_0 L\rangle,
\end{equation}
where $|i_1,i_2,...,i_l\rangle = \frac{1}{\sqrt{l!}} \sum_{\mathcal{P} \in \frak{S}_l}\text{sgn}(\mathcal{P}) |c_{\mathcal{P}i_1}, s_{\mathcal{P}i_2}, ..., s_{\mathcal{P}i_l}\rangle$.The entanglement entropy between two subsystems is then \begin{equation}
S=- {2l \choose l} \frac{(l!)^2}{(2l)!} \log \frac{l!^2}{(2l)!}.
\end{equation}
In the thermodynamic limit $l \to \infty$, this is approximated by the sterling formula as \begin{align*}
S &\simeq 2l (\log(2l) - 1) -2 l( \log l -1 ) \\
&\simeq L \log 2 . 
\end{align*}

\subsection{Entanglement entropy of integrable excited states}

An integrable eigenstate labeled by the $n$ momenta $\{\theta_1,\cdots,\theta_n\}$ of the identical color $c_1$, and the colors $\{c_2,\cdots,c_{L-n+1}\}$ of the rest $L-n$ sites can be expressed as the superposition of bipartition across the middle over different possibility of coloring in the left half system specified by the number of site in the identical color $n_A$, and the combination of the rest of the colors $C$:
\begin{equation}
	\begin{split}
	|\{\theta_1,\cdots,\theta_n\},\{c_1; c_2,\cdots,c_{L-n+1}\}\rangle= \sum_{n_A=n-l}^l\sum_{C=1}^{\binom{L-n}{l-n_A}}&\text{sgn}(C)\sqrt{\frac{(l-n_A)!(l-n+n_A)!}{(L-n)!}}\sum_{\{i_1,\cdots,i_{n_A}\}} \sum_{\{i_{n_A+1},\cdots,i_n\}}\frac{w_{i_1,\cdots,i_n}}{\sqrt{\mathcal{N}} }\\
	&\frac{1}{\sqrt{(l-n_A)!}}|i_1,\cdots,i_{n_A}\rangle\otimes \frac{1}{\sqrt{(l-n+n_A)!}}|i_{n_A+1},\cdots,i_{n}\rangle,
	\end{split}
\end{equation}
where $\mathrm{sgn}(C)$ is the signature of the permutation of the coloring of the whole system divided by the those of the two subsystems, and  $\mathcal{N}$ takes care of the normalization for the usual Bethe wave function without antisymmetric basis. Plugging in the definition of the bases vectors \eqref{eq:basis}, and taking partial trace of the density matrix over the right subsystem, we get the reduced density matrix of the left subsystem
\begin{equation}
	\begin{split}
	\rho_A= &\mathrm{tr}_B |\{\theta_1,\cdots,\theta_n\},\{c_1; c_2,\cdots,c_{L-n+1}\}\rangle\langle\{\theta_1,\cdots,\theta_n\},\{c_1; c_2,\cdots,c_{L-n+1}\}|\\
	=&\sum_{n_A=n-l}^l\sum_{C=1}^{\binom{L-n}{l-n_A}}\frac{(l-n_A)!(l-n+n_A)!}{(L-n)!}\sum_{\{i_1,\cdots,i_{n_A}\}}\frac{\sum_{\{i_{n_A+1},\cdots,i_n\}}|w_{i_1,\cdots,i_n}|^2}{\mathcal{N}}|i_1,\cdots,i_{n_A}\rangle\langle i_1,\cdots,i_{n_A}|\\
	=&\bigoplus_{n_A=n-l}^l\bigoplus_{C=1}^{\binom{L-n}{l-n_A}}\frac{(l-n_A)!(l-n+n_A)!}{(L-n)!}\rho_A^{\mathrm{XXZ}}(n_A),
	\end{split}
\end{equation}
where $\rho_A^{\mathrm{XXZ}}(n_A)$ denotes the block of the reduced density matrix of the XXZ chain with $n_A$ spin down in the left subsystem. The entanglement entropy of the integrable excited states are computed from the Schmidt coefficients
\begin{equation}
	p_A(n_A;C;\{i_{n_A+1},\cdots,i_n\})=\frac{(l-n_A)!(l-n+n_A)!}{(L-n)!}p_A^{\mathrm{XXZ}}(n_A;\{i_{n_A+1},\cdots,i_n\}),
\end{equation}
which are given in terms of the Schmidt coefficients of the corresponding XXZ excited states $p_A^{\mathrm{XXZ}}(n_A;\{i_{n_A+1},\cdots,i_n\})$.
\begin{align}
	S_A=&-\sum_{n_A=n-l}^l\sum_{\{i_1,\cdots,i_{n_A}\}}p_A^{\mathrm{XXZ}}(n_A;\{i_{n_A+1},\cdots,i_n\})\log\big( \frac{(l-n_A)!(l-n+n_A)!}{(L-n)!}p_A^{\mathrm{XXZ}}(n_A;\{i_{n_A+1},\cdots,i_n\})\big)\\
	=&S_A^{\mathrm{XXZ}}+\log(L-n)!-\sum_{n_A=n-l}^lp_A^{\mathrm{XXZ}}(n_A)\big(\log(l-n_A)!+\log(l-n+n_A)!\big),
\end{align}	
where $p_A^{\mathrm{XXZ}}(n_A)=\mathrm{tr}_A \rho_A^{\mathrm{XXZ}}(n_A)$ is the sum of the Schmidt coefficient of XXZ eigenstates with a fixed number of down spins $n_A$ in subsystem A. So the entanglement entropy of our integrable excited states factorizes nicely into the entanglement contribution from the Bethe Ansatz state and the antisymmetrized bases. In the thermodynamic limit of $L=2l\to\infty$, 
\begin{equation}
	S_A-S_A^{\mathrm{XXZ}}\sim (2l-n)\log(2l-n)-\sum_{n_A=n-l}^lp_A^{\mathrm{XXZ}}(n_A)\big((l-n_A)\log(l-n_A)+(l-n+n_A)\log(l-n+n_A)\big).
\end{equation}
 The additional entanglement entropy from the bases varies with the number of colors involved, and the Schmidt decomposition of the particular XXZ eigenstate in question. It ranges between $0$ and $(L-n)\log2$. So the entanglement entropy of the integrable excited states are lower-bounded by those of the corresponding Bethe Ansatz states in spin-$\frac{1}{2}$ model, which are area law breaking~\cite{Alba_2009,Chen_2013}, and upper-bounded by volume law, as a function of the number of components available in the Hilbert space.

\section{\label{sec:summ} Conclusions  and Outlooks}

In this paper we have addressed the integrable excited states in a partially integrable $\frak{S}_N$-invariant antiferromagnetic multi-component Heisenberg chain. The model is not integrable in symmetry sectors involving more than two components, because three-body scatterings between quasi-particles  depend generally on the ordering of the factorized two-body scatterings, as interaction between identical and different components are different. However, in addition to the more apparently integrable Krylov subspaces studied before~\cite{Kiwata:1994vs,Sato:1995we,Sato:1996vm}, we found a large number of Bethe Ansatz integrable excited states in the symmetry sectors involving more than two components, by constructing an antisymmetized basis among components that appear once in the chain. The excited states we found differ from the previously studied set in the integrable symmetry sectors in that an initial state in the non-integrable sector overlaps both with integrable eigenstates and ETH-violating ones. So there will be consequences for what concerns the slow dynamics of local observables toward thermalization as a weak form of ETH violation, i.e. a process that interpolates between integrable and chaotic systems. 

While our discussion has referred to a local Hilbert space of arbitrary $N$ dimension (we have even used thermodynamically large $N$ as an extreme case to illustrate the scaling of the contribution to entanglement entropy from the antisymmetrized bases), it is worth stressing that these integrable excited states exist even for as small a number of components as $N=3$: this implies that, employing a mapping between $SU(3)$ and $SU(2)$ spin-1, these states can be realized in cold atom experiment. Although as $N$ gets smaller, integrable excited states are more outnumbered by the non-integrable ones, we speculate that the slowdown in thermalization can still be observed with a smart choice of initial state in an array of smaller number of cold atoms. However a proposal for the realization of the anisotropy interaction goes beyond the scope of the current manuscript and is deferred to future work.

Our approach to arrive at weak ergodicity breaking should be compared with the paradigm of QMBS. In those models, one usually starts with a non-integrable model of which little is known with the possible exception of a solvable ground state. So the non-integrability is established by numerical evidence of level-spacing statistics. Then by applying SMA to the solvable ground state, one can construct a tower of exact excited states, usually with $\pi$ momenta. In our case instead, we start with a model that is Bethe Ansatz integrable, and adds deformation to break integrability, so we can directly check the YBE to see the generic breakdown of integrability, as well as identify the surviving integrable eigenstates. Although the integrable excited states have finite energy density, they do not fall into the category of exact quantum many-body scar states, in the following sense. First, given these states are solved with Bethe Ansatz, their corresponding energy are not integer or rational valued, nor are they equal distant in the spectrum. So the periodic revival on top of a slow dynamics characteristic to QMBS will not be observable here. Second, they exhibit severe violation of area law and are not expressible by matrix product states with finite bond dimension~\cite{moudgalya2020large,PhysRevB.98.235156}. 

We emphasize that violation of YBE only implies the absence of non-diffractive scattering. When Bethe Ansatz fails, Sommerfeld diffraction ansatz can still apply if internal consistency depending on solution of certain Riemann-Hilbert problem is met. Indeed, McGuire and Hurst  has developed algebraic formulation for solving the three-body problem with different scattering between particles of different species~\cite{McGuire:1972aa,McGuire:1988aa}. Yet so far, this approach has only been applied to three- or four-body problems \cite{Bibikov:2016aa}, except for scattering with bound states \cite{McGuire:1964aa}. At the end of Sec.~\ref{sec:multi}, we have identified exceptional momenta that allow such exact analytical solutions by solving the YBE as equations of momenta for arbitrary anisotropy. We believe that this method can be more widely applicable than model specific result. For instance, one can apply Bethe Ansatz instead of SMA to the frustration-free ground state of AKLT chain, and solve the YBE of scattering matrices to either find momenta different from $\pi$ that allows multi-mode exact excited state, or have a definitive explanation of the relation between QMBS and $\pi$-momenta. Another direction to pursue in terms of diagonalization without going into the most general Sommerfeld diffraction ansatz is to calculate the diffractive scattering amplitude for the $\Delta\to 0$ case, when the Yang-Baxter equation is weakly violated~\cite{PhysRevA.87.012707}. 

We have been able to express the entanglement entropy of our integrable excited state as a sum of the entanglement entropy of the corresponding Bethe Ansatz eigenstates in the XXZ chain, and an additional entanglement contribution from the antisymmetrized bases. This gives another way to see how the integrable excited states we found are more non-trivial than those in the integrable symmetry sector previously studies. We believe such a sum of entanglement entropy contributions may exist elsewhere, such as the in the isotropic multi-component model integrable by nested Bethe Ansatz. We also showed that when the number of components are large enough, the ground state have volume law scaling of entanglement entropy, and the spectrum is gapless. It would be interesting to see how these features change as the number of local degrees of freedom becomes finite, either from numerics or a combination of analytical and numerical methods.

\begin{acknowledgments}
	We thank Hosho Katsura for discussions, and especially for bringing to our attention related results in the literature, and thank Lorenzo Piroli for discussions. ZZ thanks Israel Klich, Pasquale Calabrese, Feng He, and Haiyuan Zou for fruitful discussions. ZZ acknowledges the kind hospitality of the workshop ``Randomness, Integrability and Universality'' at the Galileo Galilei Institute for Theoretical Physics during the final stage of the work. The work of ZZ was partly supported by the National Natural Science Foundation of China Grant No. 12005129, and GM's work was supported in part by the grant Prin 2017-FISI.
\end{acknowledgments}


\appendix

\section{Parameterization of Scattering matrices}
\label{sec:parametrization}
The reparameterization of momenta into rapidities for different anisotropy parameters is due to Orbach \cite{PhysRev.112.309}.
\paragraph{$\Delta>1$:}
Using parameterization $\Delta\equiv \cosh\eta$, and
\begin{equation}
\mu\equiv -\frac{\sin(\lambda +i\eta/2)}{\sin(\lambda -i\eta/2)},
\end{equation}
\eqref{eq:scatterph} becomes
\begin{equation}
\frac{A_{\sigma'}(\eta)}{A_{\sigma}(\eta)}=\frac{\sin(\lambda_{\sigma (a_0+1)}-\lambda_{\sigma a_0}+i\eta)}{\sin(\lambda_{\sigma (a_0+1)}-\lambda_{\sigma a_0}-i\eta)},
\end{equation}
\paragraph{$\Delta=1$:} In this case, we parameterize
\begin{equation}
\mu\equiv -\frac{\lambda+i/2}{\lambda-i/2},
\end{equation}
and \eqref{eq:scatterph} takes the form
\begin{equation}
\frac{A_{\sigma'}(\eta)}{A_{\sigma}(\eta)}=\frac{\lambda_{\sigma (a_0+1)}-\lambda_{\sigma_{a_0}}+i}{\lambda_{\sigma (a_0+1)}-\lambda_{\sigma_{a_0}}-i}
\end{equation}
\paragraph{$0\le\Delta<1$:}
Using parameterization $\Delta\equiv\cos\eta$, and
\begin{equation}
\mu\equiv -\frac{\sinh(\lambda +i\eta/2)}{\sinh(\lambda -i\eta/2)},
\end{equation}
\eqref{eq:scatterph} then becomes
\begin{equation}
\frac{A_{\sigma'}(\eta)}{A_{\sigma}(\eta)}=\frac{\sinh(\lambda_{\sigma(a_0+1)}-\lambda_{\sigma_{a_0}}+i\eta)}{\sinh(\lambda_{\sigma(a_0+1)}-\lambda_{\sigma_{a_0}}-i\eta)},
\end{equation}
Summarizing, we have 
\begin{equation}
\label{eq:paravp}
\varphi^{\Delta}(\lambda^\Delta \pm i\eta)=
\begin{cases}
\sin(\lambda^\Delta \pm i \arccosh \Delta), & \Delta>1,\\
\lambda^\Delta \pm i,& \Delta=1\\
\sinh(\lambda^\Delta \pm i\arccos \Delta), \quad & 0\le\Delta<1,
\end{cases}
\end{equation}
and
\begin{equation}
\label{eq:paramu}
\mu=
\begin{cases}
-\frac{\sin(\lambda^\Delta +i\eta/2)}{\sin(\lambda^\Delta -i\eta/2)}, & \Delta>1,\\
-\frac{\lambda^\Delta+i/2}{\lambda^\Delta-i/2},& \Delta=1\\
-\frac{\sinh(\lambda^\Delta +i\eta/2)}{\sinh(\lambda^\Delta -i\eta/2)}, \quad & 0\le\Delta<1.
\end{cases}
\end{equation}

\section{Normalization of eigenstates}
\label{sec:norm}
In the absence of the interaction term, the eigenvalues of translation operator are roots of unity. Therefore it admits a cute calculation of the normalization constant, as we demonstrate below.
\begin{align*}
\mathcal{N}^2 =&\sum_{1\le i_1<...<i_n\le N}|w_{i_1,...,i_n}|^2\\
=&\sum_{1\le i_1<...<i_n\le L}\sum_{\sigma,\rho \in S_n} (-1)^{\sigma+\rho}\prod_{a=1}^n(\mu_{\sigma a}\mu_{\rho a}^*)^{i_a}\\
=&\sum_{1\le i_1<...<i_n\le L}\sum_{\sigma,\tau \in S_n} (-1)^\tau \prod_{a=1}^n(\mu_{\sigma a}\mu_{\tau\sigma a}^*)^{i_a} \qquad &(\tau=\rho\sigma^{-1})\\
=&\sum_{1\le i_1<...<i_n\le L}\sum_{\sigma,\tau \in S_n}(-1)^\tau\prod_{a=1}^n(\mu_a\mu_{\tau a}^*)^{i_{\sigma^{-1} a}} \qquad &(\text{dummy index}\ a)\\
=&\sum_{1\le i_1<...<i_n\le L}\sum_{\sigma,\tau \in S_n}(-1)^\tau\prod_{a=1}^n(\mu_a\mu_{\tau a}^*)^{i_{\sigma a}}  \qquad &(\text{dummy index } \sigma)\\
=&\sum_{1\le i_1\ne...\ne i_n\le L}\sum_{\tau\in S_n}(-1)^\tau\prod_{a=1}^n(\mu_a\mu_{\tau a}^*)^{i_a}\\
=&\sum_{\tau\in S_n}(-1)^\tau\prod_{a=1}^n\sum_{i_a=1}^L(\mu_a\mu_{\tau a}^*)^{i_a}-\sum_{\substack{\exists b,c,\\ i_b=i_c}}\sum_{\tau\in S_n}(-1)^\tau (\mu_b\mu_{\tau b}^*\mu_c\mu_{\tau c}^*)^{i_b}\prod_{a\ne c,d}(\mu_a\mu_{\tau a}^*)^{i_a} \\
=&\sum_{\tau\in S_n}(-1)^\tau\prod_{a=1}^n(\delta_{a,\tau a}L) \qquad &(\text{as } \mu_a\text{'s are distinct})\\
&-\sum_{\substack{\exists b,c,\\ i_b=i_c}}\sum_{\tau\in S_n/S_2}(-1)^\tau\prod_{a\ne c,d}(\mu_a\mu_{\tau a}^*)^{i_a}\sum_{\pi\in \{id,(b\ c)\}}(-1)^\pi (\mu_b\mu_{\tau b}^*\mu_c\mu_{\tau c}^*)^{i_b}  \\
=&\sum_{\tau\in S_n}(-1)^\tau \delta_{\tau, id}L^n\\
=&L^n.\qedhere
\end{align*}
As a special case, when $n=L$, this norm can be easily calculated by representing the wave function as Slater determinant, and evaluating the product of determinants with the determinant of the product of the two matrices.

\section{Derivation of scattering matrix}
\label{sec:Smat}
In this appendix, we give the detailed derivation of Yang's scattering $Y$ matrix in the reflection(-diagonal) representation. 
Using \eqref{eq:multieig}, \eqref{eq:multisgn} and \eqref{eq:multiwf}, the eigenvalue equation $H|v\rangle=E|v\rangle$, when considering adjacent identical particles case $Qa=Q(a+1)-1$ separately, gives
\begin{equation*}
\begin{split}
&\sum_Q\biggl(\sum_{x_{Qi}+1<x_{Q(i+1)}}\psi_Q\bigl[\sum_{i=1}^n-\bigl(|...,x_i-1,...\rangle+|...,x_i+1,...\rangle\bigr)-(L-2n)|...\rangle\bigr]\\
&+\sum_{x_{Qa}+1=x_{Q(a+1)}}\psi_Q\bigl[\sum_{i\ne k,k+1}-\bigl(|...,x_{i}-1,...\rangle+|...,x_{i}+1,...\rangle\bigr)\\
&-|...,x_a-1, x_{a+1},...\rangle-|...,x_a, x_{a+1}+1,...\rangle-(L-2n+1)|...,x_{i},...\rangle\\
&-|...,x_{a+1},x_a,...\rangle+2\Delta \delta^{a,a+1}|...,x_a,x_{a+1},...\rangle\bigr]+\cdots \biggr)\\
&=\sum_Q\sum_{1\le x_{Q1}<...<x_{Qn}\le L}E \psi_Q|...\rangle,
\end{split}
\end{equation*}
where $\delta^{a,a+1}=1$ if $c_{Qa}=c_{Q(a+1)}$, and $\delta^{a,a+1}=0$ if $c_{Qa}\ne c_{Q(a+1)}$, and we have omitted the unchanged variables in the ket vectors, and the ellipsis in the sum on the l.h.s. denote terms with multiple adjacent identical particles. Collecting coefficients of the same basis vectors on both sides, we have
\begin{equation}
-\sum_{i=1}^n[\psi_Q(\mathbf{x}-\mathbf{e_i})+\psi_Q(\mathbf{x}+\mathbf{e_i})]-(L-2n)\psi_Q(\mathbf{x}) =E \psi_Q(\mathbf{x}),
\end{equation}
for the non-adjacent case, where $\mathbf{e_i}$ denotes the unit vector in the $i$'th coordinate component, and
\begin{equation}
\begin{split}
&-\sum_{i=1}^n[\psi_Q(\mathbf{x}-\mathbf{e_i})+\psi_Q(\mathbf{x}+\mathbf{e_i})]
+\psi_Q(\mathbf{x}+\mathbf{e_a})+\psi_Q(\mathbf{x}-\mathbf{e_{a+1}})-(L-2n+1)\psi_Q(\mathbf{x})\\
&-\psi_{\tau_{a(a+1)} Q}(\mathbf{x}\tau_{a(a+1)})+2\Delta \delta^{a,a+1} \psi_Q(\mathbf{x})
=E \psi_Q(\mathbf{x}),
\end{split}
\end{equation}
for the adjacent case, where $\tau_{a(a+1)}$ denotes the transposition between $a$ and $a+1$. The difference between the above two equations then gives the boundary condition
\begin{equation}
\psi_Q(\mathbf{x}+\mathbf{e_a})+\psi_Q(\mathbf{x}-\mathbf{e_{a+1}})-\psi_Q(\mathbf{x})-\psi_{\tau_{a(a+1)} Q}(\mathbf{x}\tau_{a(a+1)})
+2\Delta \delta^{a,a+1}\psi_Q(\mathbf{x})=0.
\end{equation}
Plugging in our Bethe trial wave function for $\phi_Q$, we have
\begin{equation}
\sum_{P\in S_n} [(\mu_{Pa}+\mu_{P(a+1)}^{-1}-1+\Delta p_{Qa,Q(a+1)})A_{Q,P}- A_{\tau_{a(a+1)} Q,P}] \prod_{i=1}^n\mu_{Pi}^{q_{Qi}}=0.
\end{equation}
Denoting $Pa,P(a+1)$ with $Pa=i$, $P(a+1)=j$, we have
\begin{equation}
\begin{split}
&\sum_{P\in S_n/Z_2} \biggl(\bigl[\mu_i\mu_j+1+(2\Delta \delta^{a,a+1}-1)\mu_{j}\bigr]A_{Q,P}-\mu_j A_{\tau_{a(a+1)} Q, P}\\
&+\bigl[\mu_i\mu_j+1+(2\Delta \delta^{a,a+1}-1)\mu_{i}\bigr]A_{Q,P\sigma_{ij}}-\mu_i A_{\tau_{a(a+1)} Q, P\sigma_{ij}}\biggr)(\mu_i\mu_j)^a\prod_{i\ne a,a+1}\mu_{Pi}^{x_{Qi}}=0.
\end{split}
\end{equation}
A sufficient condition for this equation to hold for any choice of $\{x_i\}$, is that the coefficients in the sum vanish term by term. Treating now $A_{Q,P}$ as components of $n!$ dimensional column vectors $\xi_P$, this becomes
\begin{equation}
\begin{split}
&\bigl(\mu_i\mu_j+1+(2\Delta \delta^{a,a+1}-1)\mu_j-\pi(\tau_{a(a+1)})\mu_{j}\bigr)\xi_P\\
+&\bigl(\mu_i\mu_j+1+(2\Delta \delta^{a,a+1}-1)\mu_i-\pi(\tau_{a(a+1)})\mu_{i}\bigr)\xi_{P\sigma_{ij}}=0,
\end{split}
\end{equation}
where $\pi$ is the \textit{left} regular representation of $S_n$, which is associated to the scattering process of a Bethe wave function~\cite{Arutyunov_2019}. From this, we can solve Yang's scattering $Y$ matrix, as defined in \eqref{eq:Ydef}, to be
\begin{equation}
Y_{ij}^{a(a+1)}(\Delta)=R_{ij}^{a(a+1)}(\Delta)\mathbb{1}+T_{ij}^{a(a+1)}(\Delta)\pi(\tau_{a(a+1)}),
\end{equation}
where the reflection coefficient
\begin{equation}
R_{ij}^{ab}(\Delta)=-\frac{(\mu_i\mu_j+1+2\Delta\delta^{ab}\mu_j)(\mu_i\mu_j+1+2(\Delta\delta^{ab}-1)\mu_i)+(\mu_i\mu_j+1)(\mu_i-\mu_j)}{(\mu_i\mu_j+1+2\Delta\delta^{ab}\mu_i)(\mu_i\mu_j+1+2(\Delta\delta^{ab}-1)\mu_i)},
\end{equation}
and transmission coefficient
\begin{equation}
T_{ij}^{ab}(\Delta)=-\frac{(\mu_i\mu_j+1)(\mu_i-\mu_j)}{(\mu_i\mu_j+1+2\Delta\delta^{ab}\mu_i)(\mu_i\mu_j+1+2(\Delta\delta^{ab}-1)\mu_i)}.
\end{equation}

\section{Nested Bethe Ansatz diagonalization for the isotropic Hamiltonian}
\label{App:nestedBA} 

The periodic boundary condition in the multiple identical species case relates wave functions in different sectors of the coordinate space\begin{equation}
	\psi_{Q}(x_1,...,\{q_j=1;c_j\},...,x_n)=\psi_{\tilde{\tau}Q}(x_1,...,\{q_j=L+1;c_j\},...,x_n),
\end{equation} where $\tilde{\tau}$ denotes the translation $(2\ 3\ \cdots\ n\ 1)$. Upon applying the ansatz wave function, this implies \begin{equation}
	A_{Q,P}=A_{\tilde{\tau}Q,\tilde{\tau}P}\mu_{P(1)}^L,
\end{equation} and consequently\begin{equation}
	\pi(\tilde{\tau})\xi_{P}=\mu_{P(1)}^L\xi_{\tilde{\tau}P},
\end{equation}since the left regular representation acts as $\pi(\tilde{\tau})A_{Q,P}=A_{\tilde{\tau}^{-1}Q,P}$. Taking $P^0=\tau_{12}\cdots\tau_{(j-1)j}$, $\varphi=\xi_{P^0}$, and using the braiding property $S_{kj}\pi(\tau_{ik})=\pi(\tau_{ik})S_{ij}$, the periodic boundary condition then leads to \begin{equation}\label{eq:eigpbc}
	\mu_j^L\varphi=S_{(j+1)j}S_{(j+2)j}\cdots S_{nj}S_{1j}S_{2j}\cdots S_{(j-1)j}\varphi,
\end{equation} for $j=1, \cdots, n$. Unlike the single identical species case, this is no longer a scalar equation, but an eigenvalue problem that requires diagonalization of the r.h.s. To proceed from here, we have to specify an irreducible representation of the permutation group $\frak{S}_n$ corresponding to sector $\{c_1^{n_1},\cdots,c_m^{n_m},c_{m+1},\cdots, c_{L-n+m}\}$ of the Hilbert space. To find the lowest energy of this sector, let's take the irrep $R=[m^{n_m},(m-1)^{n_{m-1}-n_m},...,2^{n_2-n_3},1^{n_1-n_2}]$.  Sutherland's approach of diagonalization of this irrep corresponding to the multi-row Young tableaux using Bethe-Yang Hypothesis is to first treat all the other $m-1$ species as the same, except the first one, then all the other $m-2$ species as the same, except the second one, and so on, so forth. This way, at each stage of the nesting, we are dealing with an irrep corresponding to a 2-row Young tableaux, which can be solved by recourse to the algebraic Bethe Ansatz as used in diagonalizing the spin-half Heisenberg model. Notice, however, at each stage, while the irrep of the spin part of the wave function forms a 2-row Young Tableaux, the spatial wave function must form the irrep corresponding to the 2-column tableaux $[2^{n_1},1^{n-n_1}]$ at the first stage, for instance. Yang's prescription to this difficulty is to consider instead of the eigenvalue problem of \eqref{eq:eigpbc}, the equivalent eigenvalue problem of \begin{equation}
	\mu_j^L\varphi=S'_{(j+1)j}S'_{(j+2)j}\cdots S'_{nj}S'_{1j}S'_{2j}\cdots S'_{(j-1)j}\varphi,
\end{equation}where \begin{equation} S'_{ij}=T_{ij}\mathbb{1}-R_{ij}\tilde{\pi}(\tau_{ij})=\frac{\lambda_i-\lambda_j+i\pi_{\tilde{R}}(\tau_{ij})}{\lambda_i-\lambda_j+i} \end{equation}is written in terms of the conjugate representation $\tilde{R}=[n_1,n-n_1]$. Their equivalence is manifested by the fact that $\pi_{\tilde{R}}(\tau_{ij})=-\pi_R(\tau_{ij})$. The advantage of adopting such a conjugate representation is that it admits a realization in terms of the the scattering matrix in the Heisenberg spin-$\frac{1}{2}$ problem of a length $n$ chain. So we can readily apply the results from algebraic Bethe Ansatz at each stage to write $\varphi$ as \begin{equation}
	\varphi=\sum_{\sigma\in \frak{S}_{n-n_1}} A_\sigma F(\Lambda_{\sigma 1},y_1)F(\Lambda_{\sigma 2},y_2)\cdots F(\Lambda_{\sigma (n-n_1)},y_{n-n_1}),
\end{equation} where $y_1<y_2<\cdots<y_{n-n_1}$ are the coordinates of the $n_1$ ``down spins'', and \begin{equation}
	F(\Lambda,y)=\prod_{j=1}^{y-1}\frac{\lambda_j-\Lambda+i/2}{\lambda_{j+1}-\Lambda-i/2}
\end{equation} are defined in terms of the set of unequal numbers to be solved from the set of coupled algebraic equations
\begin{align}
	\mu_j^L=&-\prod_{k=1}^{n}\frac{\lambda_j-\lambda_k-i}{\lambda_j-\lambda_k+i}
	\prod_{\alpha=1}^{n-n_1}\frac{\lambda_j-\Lambda_\alpha^{(1)}+i/2}{\lambda_j-\Lambda_\alpha^{(1)}-i/2},  &j=1,...,n, \\
	\prod_{j=1}^n\frac{\Lambda_\alpha^{(1)}-\lambda_j-i/2}{\Lambda_\alpha^{(1)}-\lambda_j+i/2}=&-\prod_{\beta=1}^{n-n_1}\frac{\Lambda_\alpha^{(1)}-\Lambda_\beta^{(1)}-i}{\Lambda_\alpha^{(1)}-\Lambda_\beta^{(1)}+i} \prod_{\gamma=1}^{n-n_1-n_2}\frac{\Lambda_\alpha^{(1)}-\Lambda_\gamma^{(2)}+i/2}{\Lambda_\alpha^{(1)}-\Lambda_\gamma^{(2)}-i/2},  &\alpha=1,...,n-n_1,\\
	\vdots& & \nonumber\\
	\prod_{\delta=1}^{n_{m-1}+n_m}\frac{\Lambda_\zeta^{(m)}-\Lambda_\delta^{(m-1)}-i/2}{\Lambda_\zeta^{(m)}-\Lambda_{\delta}^{(m-1)}+i/2}=&-\prod_{\epsilon=1}^{n_m}\frac{\Lambda_\zeta^{(m)}-\Lambda_\epsilon^{(m)}-i}{\Lambda_\zeta^{(m)}-\Lambda_{\epsilon}^{(m)}+i}, & \zeta=1,...,n_m.
\end{align} This set of equations are historically called Lieb-Wu equations \cite{PhysRevLett.20.1445}, the details of their derivation of these equations can be found in standard texts such as \cite{Arutyunov_2019}.
Now if there's a cut-off on the number of components, the ground state of our Hamiltonian will no longer be a superposition of fully antisymmetrized spin configurations. In stead, from the reasoning above, it will be in the sector where each color appear the same number of times along the chain, say the number of colors $s$ divides $L$, and $l=L/s$. In this sector, the Young tableaux corresponding to the lowest energy eigenstate in this sector will be $[l^s]$. The Lieb-Wu equations for this irrep become \begin{align}
	L\theta(\lambda_j)=&2\pi J_j^{(0)}+\sum_{k=1}^{L}\theta\big(\frac{1}{2}(\lambda_k-\lambda_j\big)
	+\sum_{\alpha=1}^{(s-1)l}\theta(\lambda_j-\Lambda^{(1)}_{\alpha}), &j=1,...,L,\\
	\sum_{j=1}^L\theta(\lambda_j-\Lambda^{(1)}_{\alpha})=&2\pi J_\alpha^{(1)}+ \sum_{\beta=1}^{(s-1)l}\theta\big(\frac{1}{2}(\Lambda^{(1)}_\beta-\Lambda^{(1)}_\alpha)\big)+\sum_{\gamma=1}^{(s-2)l}\theta(\Lambda^{(1)}_\alpha-\Lambda^{(2)}_\gamma), &\alpha=1,...,(s-1)l,\\
	\vdots& &\nonumber\\
	\sum_{\delta=1}^{2l}\theta(\Lambda_{\delta}^{(s-1)}-\Lambda_{\zeta}^{(s)})=& 2\pi J^{(s)}_\zeta+\sum_{\epsilon=1}^{l}\theta\big(\frac{1}{2}(\Lambda^{(s)}_\epsilon-\Lambda^{(s)}_\zeta)\big), &\zeta=1,...,l,
\end{align}where $\theta(\lambda)=2\cot^{-1}(2\lambda)$, and the quantum numbers that labels the eigenstates, $J$'s, are half-integers.

\bibliography{AFMXXZPRB2nd}

\end{document}